\documentclass[aps,prd,twocolumn,groupedaddress,nofootinbib]{revtex4}
\usepackage{graphics}

\def\Slash#1{\hskip 0.05 cm \slash\hskip -0.26 cm #1} %Feynmann slash notation
\usepackage{latexsym}

\begin{document}
\title{Reheating of the universe after inflation with $f(\phi)R$ gravity}
\author{Yuki Watanabe}
\email{yuki@astro.as.utexas.edu}
\affiliation{Department of Physics, University of Texas, Austin, Texas
78712, USA} 
\author{Eiichiro Komatsu}
\affiliation{Department of Astronomy, University of Texas, Austin, Texas
78712, USA} 
\date{\today}
\begin{abstract}
 We show that reheating of the universe occurs spontaneously in a broad
 class of inflation models with $f(\phi)R$ gravity ($\phi$ is
 inflaton). The model does not require explicit couplings between
 $\phi$ and bosonic or fermionic matter fields. The couplings arise
 spontaneously when $\phi$ settles in the vacuum expectation value
 (vev) and oscillates, with coupling constants given by 
 derivatives of $f(\phi)$ at the vev and the mass of resulting bosonic or
 fermionic fields. This mechanism allows inflaton quanta to decay into any
 fields which are not conformally invariant in $f(\phi)R$ gravity theories.
\end{abstract}

\maketitle

Inflation is an indispensable building-block of the standard model of
cosmology \cite{Linde, KolbTurnerLiddleLyth}, and has passed a number
of stringent observational tests \cite{WMAP1WMAP3}. Any inflation models
must contain a mechanism by which the universe reheats after
inflation \cite{reheating}. The
reheating mechanism requires detailed knowledge of 
interactions (e.g., Yukawa coupling) between inflaton fields and their
decay products. Since the physics behind inflation is beyond the
standard model of 
elementary particles, the precise nature of inflaton fields is currently
undetermined, and the coupling between inflaton and matter fields is
often put in by hand.
On the other hand, inflaton and matter fields are coupled through
gravity. Of course the gravitational coupling is suppressed by the
Planck mass and thus too weak to yield interesting effects \cite{Linde}; however, 
we shall show that the reheating occurs spontaneously if inflaton is coupled
to gravity non-minimally, i.e., the gravitational action is given not by
the Einstein-Hilbert form, $R$, but by $f(\phi)R$, where $\phi$ is
an inflaton field. Even if matter fields do not interact with $\phi$
directly, they ought to interact via gravitation whose perturbations
lead to Yukawa-type interactions. A similar idea was put forward by 
\cite{BassettLiberati98}, who considered {\it preheating} with a
non-minimal coupling between matter and gravity.
Here, we do not consider preheating, but focus only on the perturbative
reheating arising from $f(\phi)R$ gravity. Therefore, we can 
%analytically 
calculate the resulting reheating temperature after inflation. 

Why study $f(\phi)R$ gravity? There are a number of
motivations \cite{ScalarTensor, Faraoni}. The strongest motivation comes
from the fact that almost any candidate theories of fundamental physics
which involve compactification of extra dimensions yield $f(\phi)R$ with
the form of $f(\phi)$ depending on models. One illuminating example
would be string moduli with $f(\phi)\propto e^{-\alpha\phi}$.
Zee's induced gravity theory \cite{Zee79} has $f(\phi)=\xi\phi^2$, and
renormalization in the 
curved spacetime yields other more
complicated higher derivative terms \cite{QFTinC}. 
Classic scalar-tensor theories, originally motivated by 
Mach's principle \cite{Dicke62}, also fall into this category. 
It has been shown that inflation occurs naturally in these generalized
gravity models \cite{extended_inflation,FutamaseMaeda89,SalopekBondBardeen89,BerkinMaeda91,FakirUnruh90FHU92,kalara/etal:1990}, 
and the spectrum of scalar curvature perturbations 
\cite{MakinoSasaki91:Kaiser95:Faraoni00} as well as of tensor gravity
wave perturbations \cite{nonminimal_coupling} can be
affected by the presence of $f(\phi)R$, thereby allowing us to constrain
$f(\phi)$ from the cosmological data. We show that limits on the
reheating temperature (e.g., gravitino problem) provide additional,
totally independent constraints on $f(\phi)$. This argument opens up 
new possibilities that one can constrain a broad class of inflation
models from the reheating temperature. 
Reheating in Starobinsky's $R^2$ inflation has been considered by \cite{vilenkin/mijic/etal,kalara/etal:1990}.

Interactions between $\phi$ and matter stem from a mixing
between metric and scalar field perturbations through $f(\phi)R$
\cite{ScalarTensor}. It is this ``gravitational decay channel'' from $\phi$ to
matter that makes the universe reheat after inflation. We realize
this in a simple Lagrangian given by
\begin{eqnarray}
\label{eq:Lagrangian}
\mathcal{L}=\sqrt{-g}\left[\frac{1}{2}f(\phi)R-\frac{1}{2}g^{\mu\nu}\partial_{\mu}\phi\partial_{\nu}\phi-V(\phi)\right]+\mathcal{L}_{\rm m},
\end{eqnarray}
where $\mathcal{L}_{\rm m}$ is the matter Lagrangian.
Note that there
is no explicit coupling between $\phi$ (inflaton) and $\mathcal{L}_{\rm
m}$. To guarantee the ordinary Einstein gravity at low energy, we
must have $f(v)=M_{\rm Pl}^2$, where $v$ is the vacuum expectation value
(vev) of $\phi$ at the end of inflation and $M_{\rm Pl}=(8\pi
G)^{-1/2}=2.436\times 10^{18}$~GeV is the reduced Planck mass.
%We work with (+++) conventions of \cite{MTW}. 
This Lagrangian satisfies  the weak equivalence principle: 
$\phi$ does not couple to matter directly, 
$\nabla_{\mu}T^{\mu\nu}_{\rm m}=0$. This, however, does not imply
that inflaton {\it quanta}, i.e., fluctuations around the vev,
cannot decay into matter.
The gravitational field equation from Eq.~(\ref{eq:Lagrangian}) is
\begin{eqnarray}
f(\phi)G_{\mu\nu}&=&T_{\mu\nu}^{\rm m}+\partial_{\mu}\phi\partial_{\nu}\phi-\frac{1}{2}g_{\mu\nu}(\partial\phi)^2-g_{\mu\nu}V(\phi) \nonumber\\
& &-(g_{\mu\nu}\Box-\nabla_{\mu}\nabla_{\nu})f(\phi)\label{eq:Einstein}.
\end{eqnarray}
Before reheating completes the energy density of the universe was
dominated by $\phi$; thus, we treat $T_{\mu\nu}^{\rm m}$ as perturbations.
As usual we decompose $g_{\mu\nu}$ into the background,
$\bar{g}_{\mu\nu}$, and perturbations, $h_{\mu\nu}$, as
$g_{\mu\nu}=\bar{g}_{\mu\nu}+h_{\mu\nu}$. The first-order perturbation
in $G_{\mu\nu}$ is given by
\begin{eqnarray}
\delta G_{\mu\nu}=\frac{1}{2}[-\Box h_{\mu\nu}+\nabla_{\lambda}\nabla_{\mu}h^{\lambda}{}_{\nu}+\nabla_{\lambda}\nabla_{\nu}h^{\lambda}{}_{\mu}
- \nabla_{\mu}\nabla_{\nu}h \nonumber\\
-\bar{g}_{\mu\nu}(\nabla_{\rho}\nabla_{\sigma}h^{\rho\sigma}-\Box h-\bar{R}_{\rho\sigma}h^{\rho\sigma})-\bar{R}h_{\mu\nu}],\label{eq:G}
\end{eqnarray}
where $h^{\mu}{}_{\nu}\equiv\bar{g}^{\mu\lambda}h_{\lambda\nu}$ and
$h\equiv\bar{g}^{\mu\nu}h_{\mu\nu}$. 

Reheating occurs at the potential minimum of $\phi$ where $\phi$
oscillates about $v$. We thus expand $\phi$ as $\phi=v+\sigma$,
where $\sigma$ represents inflaton quanta which decay into matter fields.
Treating $\sigma$ as perturbations, 
one obtains the linearized field equation:
\begin{eqnarray}
\frac{M_{\rm Pl}^2}{2}\left[-\Box h_{\mu\nu}+\cdots\right]+f'(v)(\bar{g}_{\mu\nu}\Box-\nabla_{\mu}\nabla_{\nu})\sigma
=T_{\mu\nu}^{\rm m},\label{eq:linear}
\end{eqnarray}
where $f'(v)$ means $\partial f/\partial\phi|_{\phi=v}$.
This equation contains both $\Box h_{\mu\nu}$ and $\Box\sigma$, and thus
wave modes are mixed up together. To diagonalize the wave mode
 we define a new field as \cite{ScalarTensor}
\begin{eqnarray}
\tilde{h}_{\mu\nu}&\equiv&
 h_{\mu\nu}-\frac{1}{2}\bar{g}_{\mu\nu}h-\frac{f'(v)}{M_{\rm
 Pl}^2}\bar{g}_{\mu\nu}\sigma,\label{eq:mixing}
\end{eqnarray}
where the tilde denotes the operation defined by this equation, which essentially corresponds to the infinitesimal conformal transformation. The inverse operation is
\begin{eqnarray}
h_{\mu\nu}=\tilde{h}_{\mu\nu}-\frac{1}{2}\bar{g}_{\mu\nu}\tilde{h}-\frac{f'(v)}{M_{\rm Pl}^2}\bar{g}_{\mu\nu}\sigma.
\label{eq:mixing2}
\end{eqnarray}
Using the harmonic gauge (analog of the Lorenz gauge) conditions,
$\nabla_{\lambda}\tilde{h}^{\lambda}{}_{\nu}=0$, one can show that
$\tilde{h}_{\mu\nu}$ indeed obeys the linearized field equation for the wave
mode (see, e.g., Eq.~(35.64) in \cite{MTW}; Eq.~(10.9.4) in \cite{Weinberg})
\begin{eqnarray}
\Box\tilde{h}_{\mu\nu}-2\bar{R}_{\mu\lambda\rho\nu}\tilde{h}^{\lambda\rho}-2\bar{R}_{\rho(\mu}\tilde{h}^{\rho}{}_{\nu)}
-\bar{g}_{\mu\nu}\bar{R}_{\lambda\rho}\tilde{h}^{\lambda\rho}+\bar{R}\tilde{h}_{\mu\nu}\nonumber\\
=-\frac{2}{M_{\rm Pl}^2}T_{\mu\nu}^{\rm m}.\qquad\label{eq:wave_curved}
\end{eqnarray}
One may also impose the
traceless condition on $\tilde{h}_{\mu\nu}$ to make sure that $\tilde{h}_{\mu\nu}$
describes tensor (spin two) gravity waves, 
$\tilde{h}=-h-4f'(v)\sigma/M_{\rm Pl}^2=0$, which relates a trace part of 
the original metric perturbations to $\sigma$ as
$h=-4f'(v)\sigma/M_{\rm Pl}^2$. In this sense $\sigma$ describes the
``scalar (spin zero)
gravity waves'', which are common in scalar-tensor theories of gravity.
For generality we keep $\tilde{h}$ explicitly throughout the paper.

We consider both fermionic, $\psi$, and bosonic, $\chi$, fields as
matter: $\cal{L}_{\rm m}={\cal L}_\psi+{\cal L}_\chi +{\cal L}_{\rm int}$. 
First, let us write ${\cal L}_\psi$ as
\begin{eqnarray}
\label{eq:spinorlagrangian}
\mathcal{L}_{\psi}&=&-\sqrt{-g}\bar{\psi}\left[\Slash{D}+m_{\psi}
					 \right]\psi,
\end{eqnarray}
where $\Slash{D}\equiv e^{\mu\alpha}\gamma_{\alpha} D_{\mu}$, 
$e^{\mu\alpha}$ is a tetrad (vierbein) field, and 
$D_{\mu}\equiv \partial_{\mu}-\Gamma_{\mu}$ is the covariant derivative 
for spinor fields \cite{Weinberg, QFTinC}.
$\Gamma_{\mu}$ is a spinor connection and $\Sigma^{\alpha\beta}$
are generators of the Lorentz group given by 
$\Gamma_{\mu}(x)\equiv
-\frac{1}{2}\Sigma^{\alpha\beta}e_{\alpha}{}^{\lambda}\nabla_{\mu}e_{\beta\lambda}$ and 
$\Sigma^{\alpha\beta}=-\Sigma^{\beta\alpha}=\frac{1}{4}[\gamma^{\alpha},\gamma^{\beta}]$, respectively. 
Here $\alpha, \beta, \dots$ denote
Lorentz indices while $\mu, \nu, \dots$ denote general coordinate
indices. 
Note that we have not antisymmetrized the first term in
Eq.~(\ref{eq:spinorlagrangian}) to make the expression
simpler. The antisymmetrized term yields the same result.
Note also that we shall ignore the existence of torsion. 

We expand the tetrad and metric into the background and perturbations as
 $e^{\mu\alpha}\simeq
 \bar{e}^{\mu\alpha}-\bar{e}^{\lambda\alpha}h^{\mu}{}_{\lambda}/2$,
and
 $\sqrt{-g}=\bar{e}+\delta e\simeq
 \bar{e}\left(1+\bar{g}^{\mu\nu}h_{\mu\nu}/2\right)$,
 respectively. 
The Lagrangian becomes (with Eq.~(\ref{eq:mixing2}))
\begin{eqnarray}
\mathcal{L}_{\psi}&\simeq&
-\bar{e}\bar{\psi}\left[\bar{e}^{\mu\alpha}\gamma_\alpha D_{\mu}+m_{\psi} \right]\psi 
\nonumber\\
& &+\frac{1}{2}\bar{e}\bar{\psi}\big[\bar{e}^{\nu\alpha}\gamma_{\alpha}(\tilde{h}^{\mu}{}_{\nu}
+\frac{1}{2}\tilde{h} \delta^{\mu}{}_{\nu})D_{\mu}+\tilde{h} m_{\psi}\big]\psi \nonumber\\
& & 
+\bar{e}g_\psi
\sigma\bar{\psi}\psi.\label{eq:spinorlagrangian2}
\end{eqnarray}
The terms that are proportional to $\tilde{h}$ may be set to
vanish by gauge transformation.
Here we have used the background Dirac equation,
$\bar{e}^{\mu \alpha}\gamma_\alpha D_\mu \psi=-m_{\psi}\psi$, to
obtain the last term and ignored the second order
terms as well as thermal mass induced by quantum corrections in thermal
bath \cite{KolbNotariRiotto03,Yokoyama06}. 
The last term in Eq.~(\ref{eq:spinorlagrangian2}) is a Yukawa
interaction term with a coupling constant given by
\begin{eqnarray}
\label{eq:Yukawa}
g_{\psi}\equiv \frac{f'(v)m_{\psi}}{2M_{\rm{Pl}}^2}.
\end{eqnarray}
Therefore, $\sigma$ can decay into $\psi$ and $\bar{\psi}$.
Note that the Yukawa interaction vanishes when $f'(v)=0$ or
$m_\psi=0$, which is consistent with previous work \cite{SalopekBondBardeen89}:
as massless fermions are conformally invariant, no peculiar effects
can be caused by $f(\phi)R$ gravity at the tree level.
Quantum corrections such as a conformal, or trace, anomaly might open
additional decay channels. 

Next, let us consider the bosonic matter, $\chi$, given by
\begin{eqnarray}
\mathcal{L}_{\chi}=\sqrt{-g}\left[-\frac{1}{2}g^{\mu\nu}\partial_{\mu}\chi\partial_{\nu}\chi-U(\chi)\right],\label{eq:lagrangianchi}
\end{eqnarray}
which may be expanded as
\begin{eqnarray}
\mathcal{L}_{\chi}\simeq \sqrt{-\bar{g}}\bigg[-\frac{1}{2}\bar{g}^{\mu\nu}\partial_{\mu}\chi\partial_{\nu}\chi-U(\chi)+\frac{1}{2}\tilde{h}^{\mu\nu}\partial_{\mu}\chi\partial_{\nu}\chi
\nonumber\\
+\frac{1}{2}\tilde{h} U(\chi)+\frac{f'(v)}{2M_{\rm Pl}^2}
\left(
4\sigma U(\chi)+\sigma\bar{g}^{\mu\nu}\partial_{\mu}\chi\partial_{\nu}\chi\right)\bigg],\label{eq:lagrangianchi2}
\end{eqnarray}
where we have used $\sqrt{-g}\simeq\sqrt{-\bar{g}}(1-\tilde{h}/2-2\sigma f'(v)/M^2_{\rm Pl})$ and $g^{\mu\nu}\simeq \bar{g}^{\mu\nu}-\tilde{h}^{\mu\nu}+\bar{g}^{\mu\nu}\tilde{h}/2+\bar{g}^{\mu\nu}\sigma f'(v)/M_{\rm Pl}^2$.
Again, the terms that are proportional to $\tilde{h}$ may be set to
vanish by gauge transformation.
For simplicity we assume that $\chi$ is a massive field with self-interaction, 
$U(\chi)= m_{\chi}^2\chi^2/2+\lambda\chi^4/4$. 
The last term in Eq.~(\ref{eq:lagrangianchi2}) then yields the 
following interactions:
$\sqrt{-{\bar{g}}}[{\sigma f'(v)}/{2M_{\rm Pl}^2}](
2m_\chi^2\chi^2+\lambda\chi^4
+\bar{g}^{\mu\nu}\partial_{\mu}\chi\partial_{\nu}\chi)$.
We rewrite these interactions as
\begin{eqnarray}
\sqrt{-{\bar{g}}}\frac{\sigma f'(v)}{2M_{\rm Pl}^2}
\left(
m_{\chi}^2\chi^2
+\bar{g}^{\mu\nu}\bar{\nabla}_{\mu}
(\chi\partial_\nu\chi)\right)\approx 
\sqrt{-{\bar{g}}}g_{\chi}\sigma\chi^2
,\label{eq:threelegged} 
\end{eqnarray}
where we have defined a coupling constant,
\begin{eqnarray}
g_{\chi} &\equiv& \frac{f'(v)}{2M_{\rm Pl}^2}\left(m_{\chi}^2+\frac{m_{\sigma}^2}{2}\right),\label{eq:g_chi} 
\end{eqnarray}
which will give a rate of $\sigma$ decaying into two $\chi$s.
To derive Eq.~(\ref{eq:threelegged}) we used
$\bar{g}^{\mu\nu}\partial_{\mu}\chi\partial_{\nu}\chi=-\chi U'(\chi)
+\bar{g}^{\mu\nu}\bar{\nabla}_\mu (\chi\partial_\nu\chi)$,
\label{eq:simple1}
where $\bar{\nabla}_\mu$ is the covariant derivative on the background metric, 
and estimated as $\sigma\bar{\nabla}_{\mu}(\chi\partial^{\mu}\chi)\approx \sigma m_{\sigma}^2\chi^2/2$.
This estimation is valid in the zero temperature limit (see
\cite{Yokoyama06b} for the high temperature limit).
While the first term in Eq.~(\ref{eq:g_chi})  is absent when
$m_\chi=0$, $\sigma$ can still decay through the second term 
as minimally coupled massless scalar fields are not conformally invariant.
%Note that $\sigma$ is not coupled to $\lambda \chi^4$ term, as $\lambda \chi^4$ is conformally invariant.

What is the physical interpretation of these couplings?
Through the field mixing that can be seen in Eq.~(\ref{eq:mixing}) the inflaton quanta
are coupled to the trace (spin zero) part of $h_{\mu\nu}$,
which describes the scalar gravity waves.
The scalar waves are then coupled to the matter fields.
If inflaton quanta are at least twice as heavy as $\psi$
or $\chi$, they decay into $\psi\bar{\psi}$ or two $\chi$s.

In order to compute decay rates, one must canonically normalize fields.\footnote{We are grateful to N. Kaloper for pointing out importance of the normalization of $\sigma$ fields.} 
The kinetic term of $\sigma$ is given by $\mathcal{L}_{\sigma}^{\rm kin}/\sqrt{-\bar{g}}=-\frac12(\partial\sigma )^2\left[1+\frac{3}{2}(f'(v)/M_{\rm Pl})^2\right]$ to the leading order, and thus $\hat{\sigma}\equiv \sigma\sqrt{1+\frac{3}{2}(f'(v)/M_{\rm Pl})^2}$ is the canonically normalized inflaton field. Therefore, coupling constants should be replaced accordingly by 
$\hat{g}_{\psi}=g_{\psi}\left[1+\frac{3}{2}(f'(v)/M_{\rm Pl})^2\right]^{-1/2}$ and 
$\hat{g}_{\chi}=g_{\chi}\left[1+\frac{3}{2}(f'(v)/M_{\rm Pl})^2\right]^{-1/2}$. 

The decay rates of $\sigma\rightarrow \psi\bar{\psi}$ from the last term in Eq.(\ref{eq:spinorlagrangian2}) and
$\sigma\rightarrow\chi\chi$ from Eq.(\ref{eq:threelegged}), 
$\Gamma_{\rm
tot}=\Gamma_{\sigma\bar{\psi}\psi}+\Gamma_{\sigma\chi\chi}+\ldots$,
are given by \cite{Yokoyama04}
\begin{eqnarray}
\Gamma_{\sigma\bar{\psi}\psi}&=&\frac{\hat{g}_{\psi}^2m_{\sigma}}{8\pi}
\left(1-\frac{4m_\psi^2}{m_\sigma^2}\right)^{3/2}
\tanh\left(\frac{m_\sigma}{4T}\right)
C_\psi, \label{eq:decayf}\\
\Gamma_{\sigma\chi\chi}&=&\frac{\hat{g}_{\chi}^2}{8\pi m_{\sigma}}
\left(1-\frac{4m_\chi^2}{m_\sigma^2}\right)^{1/2}
\coth\left(\frac{m_\sigma}{4T}\right)
C_\chi,
\label{eq:decay}
\end{eqnarray} 
where $m_{\sigma}^2\equiv V''(v)$ and
we have included suppression of decay rates due to thermal mass,
$C_\psi$ and $C_\chi$, which are unity in the low temperature limit
($T\ll m_\sigma$) in which thermal mass is small, but can be quite small
otherwise. 
The exact form of $C$ depends on how decay products are thermalized.
Yokoyama has calculated $C_\chi$ from thermalization due to $\Delta
U=\lambda\chi^4/4$: for $T\gg m_\sigma/\sqrt{\lambda}$ he finds
$C_\chi=\lambda/(8\pi^2)$ for decay via $\sigma\chi^2$ interaction 
\cite{Yokoyama06} and
$C_\chi=\lambda/(1024\pi^2)$ for  $\sigma(\partial \chi)^2$
(which corresponds to the last term in Eq.~(\ref{eq:lagrangianchi2}); if
this term dominates $g_\chi$ is given by $g_\chi=f'(v)\lambda T^2/(2M_{\rm
Pl}^2)$ because of dominance of thermal mass, $\sqrt{\lambda}T$, over
the intrinsic mass of $\chi$) 
\cite{Yokoyama06b}, while $C_\psi$ has not been calculated explicitly yet.

As usual we use the condition, $3H(t^*)=\Gamma_{\rm tot}$, to define the
reheating time, $t^*$, at which radiation begins to dominate the energy
density of the universe. From the Friedmann equation, $H^2=\rho/(3M_{\rm
Pl}^2)$, we get $\rho(t^*)=\Gamma_{\rm tot}^2M_{\rm Pl}^2/3=\frac{\pi^2}{30}g_*(T_{\rm rh})T_{\rm rh}^4$,
where the last equality holds if all the decay products interact with each
other rapidly enough to achieve thermodynamical equilibrium \cite{KolbNotariRiotto03}. 
If some fraction of inflaton energy density is converted into the species that never interact with the visible sector, the reheating temperature may be lowered. 
$g_*(T_{\rm rh})$ is the effective number of relativistic degrees of freedom at the
reheating time. 
Since $\Gamma_{\rm
tot}>\Gamma_{\sigma\bar{\psi}\psi}+\Gamma_{\sigma\chi\chi}$, the reheating
temperature is bounded from below: 
\begin{eqnarray}
T_{\rm rh} &=& \frac{\sqrt{\Gamma_{\rm tot}M_{\rm Pl}} }{(10\pi^2)^{1/4}}\left(\frac{g_*(T_{\rm rh})}{100}\right)^{-1/4}\nonumber\\
&>&
 \sqrt{\hat{g}_{\psi}^2+\frac{\hat{g}_{\chi}^2}{m_{\sigma}^2}}\frac{\sqrt{m_{\sigma}M_{\rm
 Pl}}}{4\pi(40)^{1/4}}\left(\frac{g_*(T_{\rm rh})}{100}\right)^{-1/4},  
\label{eq:trh}
\end{eqnarray}
where we have used Eq.~(\ref{eq:decayf}) and (\ref{eq:decay}) assuming
for simplicity that decay products are much lighter than $\sigma$
($m_\psi$, $m_\chi\ll m_\sigma$), 
the reheating temperature is much smaller than $m_\sigma$
($T\ll m_\sigma$), and therefore suppression due to thermal mass is unimportant 
($C_\psi=1=C_\chi$). Generalization to the other limits is straightforward.
One may reverse the argument
and set a limit on $f'(v)$ as
\begin{eqnarray}
\frac{|f'(v)|}{\sqrt{1+\frac{3}{2}\left(\frac{f'(v)}{M_{\rm Pl}}\right)^2}}< 8\pi(40)^{1/4} T_{\rm rh}\left(\frac{M_{\rm Pl}}{m_{\sigma}}\right)^{3/2}\left(\frac{g_*(T_{\rm rh})}{100}\right)^{1/4},
\end{eqnarray}
which provides non-trivial constraints on inflation models with
$f(\phi)R$ gravity. 
Let us consider a non-minimal coupling model, $f(\phi)=M^2+\xi\phi^2$,
for instance ($M^2$ is given by the condition $f(v)=M_{\rm Pl}^2$).
We obtain
$|\xi|(1+6\xi^2 v^2/M_{\rm Pl}^2)^{-1/2} 
< 4\pi(40)^{1/4} (T_{\rm rh}/v)(M_{\rm Pl}/m_{\sigma})^{3/2}
(g_*(T_{\rm rh})/100)^{1/4}$,
which provides a new constraint on 
$\xi$, independent of the existing constraints from homogeneity and isotropy
\cite{FutamaseMaeda89} and curvature as
well as tensor perturbations
\cite{nonminimal_coupling}.  

Finally, let us show that 
one can obtain the same results
by performing the following conformal transformation,
$\hat{g}_{\mu\nu}=\Omega^2 g_{\mu\nu}$, with
\begin{eqnarray} 
\Omega^2&=&\frac{f(\phi)}{M_{\rm Pl}^2}
= 1+\frac{f'(v)\sigma}{M_{\rm Pl}^2}
+\frac{f''(v)\sigma^2}{2M_{\rm Pl}^2}
+{\cal O}(\sigma^3).
\label{eq:conformal_tr}
\end{eqnarray} 
Hereafter we put hats on variables in the Einstein frame (E frame). 
We use Maeda's transformation formula \cite{Maeda89} to obtain
the Lagrangian in the E frame,
\begin{eqnarray}
\mathcal L=\sqrt{-{\hat g}}\bigg[\frac{M_{\rm Pl}^2}{2}{\hat R}-\frac{1}{2}{\hat g}^{\mu\nu}\partial_{\mu}\hat{\phi}\partial_{\nu}\hat{\phi}-{\hat V}(\hat{\phi})\bigg]
+{\mathcal L}_{\rm m},
\end{eqnarray}
where $\hat{R}$ is calculated from $\hat{g}_{\mu\nu}$, a
transformed
potential is given by
$\hat{V}(\hat{\phi})\equiv \Omega^{-4}V(\phi)$, and
a new scalar field, $\hat{\phi}$, is defined such that it has the
canonical kinetic term. It is related to the original field by 
 $d\hat{\phi}/d\phi = 
M_{\rm Pl}\sqrt{1/f(\phi)+\frac{3}{2}\left(f'(\phi)/f(\phi)\right)^2}$.
As $\hat{\phi}$ is minimally coupled to $\hat{R}$, there is no mixing between
$\hat{g}_{\mu\nu}$ and $\hat{\phi}$.

A transformed Lagrangian for $\psi$ in the E frame is
\begin{eqnarray}\label{eq:matterlagrangian_E}
{\mathcal L}_{\psi}=-\hat{e}\hat{\bar \psi}[(\hat{\Slash{D}}-\hat{\overleftarrow{\Slash{D}}})/2+\Omega^{-1}m_{\psi}]\hat{\psi},
\end{eqnarray}
where 
$\hat{\psi}=\Omega^{-3/2}\psi,\; \hat{\bar
\psi}=\Omega^{-3/2}{\bar \psi}$, 
$\hat{e}^{\mu\alpha}=\Omega^{-1}e^{\mu\alpha}$, and $\hat{e}=\Omega^{4}e$. 
While no Yukawa interaction arises from the
kinetic term, the  mass term yields a Yukawa interaction:
\begin{eqnarray}
-\hat{e}\frac{m_{\psi}}{\Omega}\hat{\bar \psi}\hat{\psi}\simeq -\hat{e}m_{\psi}\hat{{\bar \psi}}\hat{\psi}+\hat{e}\frac{f'(v)m_{\psi}}{2M_{\rm Pl}^2}\sigma\hat{\bar \psi}\hat{\psi}.
\end{eqnarray}
Thus, Yukawa coupling constants agree precisely.

A transformed Lagrangian for $\chi$ in the E frame is
\begin{eqnarray}
\mathcal{L}_{\chi}= -\frac{1}{2}\sqrt{-\hat{g}}\hat{g}^{\mu\nu}\mathcal{D}_{\mu}\hat{\chi}\mathcal{D}_{\nu}\hat{\chi}-\sqrt{-\hat{g}}\hat{U}(\hat{\chi}),\label{eq:lagrangianchi_E}
\end{eqnarray}
where $\hat{\chi}\equiv\Omega^{-1}\chi$ and we have followed the
procedure in \cite{ScalarTensor} to define a ``covariant derivative'' 
and potential in the E frame as
$\mathcal{D}_{\mu}\equiv
\partial_{\mu}+\partial_{\mu}(\ln\Omega)$,
and $\hat{U}(\hat{\chi})\equiv \Omega^{-4}U(\chi)$, respectively.
In the case of $U(\chi)=m_{\chi}^2\chi^2/2+\lambda\chi^4/4$, the second term of
Eq.~(\ref{eq:lagrangianchi_E}) becomes  
\begin{eqnarray}
-\sqrt{-\hat{g}}\left[\frac{1}{2}m_{\chi}^2\hat{\chi}^2+\frac{\lambda}{4}\hat{\chi}^4-\frac{f'(v)m_{\chi}^2}{2M_{\rm Pl}^2}\sigma\hat{\chi}^2\right],
\end{eqnarray}
to the linear order in $\sigma$; the last term agrees with the first term in Eq.~(\ref{eq:threelegged}).
The covariant derivative yields a coupling,
$-\sqrt{-\hat{g}}\hat{\chi}\hat{g}^{\mu\nu}(\partial_\mu\hat{\chi})(\partial_\nu\ln \Omega)$.
After integration by parts and $\ln\Omega\simeq [f'(v)/(2M_{\rm
Pl}^2)]\sigma$ this term yields
\begin{equation}
 \frac{f'(v)}{2M_{\rm  Pl}^2}\sigma\partial_\nu(\sqrt{-\hat{g}}\hat{\chi}\hat{g}^{\mu\nu}\partial_\mu\hat{\chi})
= \sqrt{-\hat{g}}\frac{f'(v)}{2M_{\rm  Pl}^2}\sigma
\hat{g}^{\mu\nu}\hat{\nabla}_\mu(\hat{\chi}\partial_\nu\hat{\chi}),
\end{equation}
which agrees with the second term in
Eq.~(\ref{eq:threelegged}) precisely.

One can calculate the other interaction terms such as 
$\sigma^2\chi^2$, $\sigma^2\bar{\psi}\psi$, $\sigma^2(\partial\chi)^2$,
etc., with known coupling constants 
in the E frame easily.
These interactions, whose coupling constants
are proportional to the
higher order derivatives such as $f''(v)$, $f'''(v)$, etc.,
actually dominate if $f(\phi)$ also has a minimum at the vev, $f'(v)=0$.
We must canonically normalize \textit{quanta} as $\hat{\sigma}=\sigma\sqrt{1+\frac32(f'(v)/M_{\rm Pl})^2}$ when calculating their interaction rates.
While we have ignored a parametric resonance (preheating)
\cite{preheating} 
entirely, it would be interesting to study how preheating might occur in
the present context.

In summary, we have presented a natural mechanism for reheating of the
universe after inflation without introducing any 
explicit couplings between inflaton and fermionic or bosonic
matter fields.  
This mechanism allows inflaton quanta to decay 
into {\it any fields} which are
present at the end of inflation and are not conformally invariant,
when inflaton settles in the vacuum expectation value and oscillates.
Reheating therefore occurs spontaneously in {\it any theories} of
$f(\phi)R$ gravity.

We have calculated the reheating temperature from this mechanism.
We argue that one must always check that the reheating
temperature in any 
$f(\phi)R$ inflation models is
reasonable, e.g., the reheating temperature does not exceed the critical
temperature above which too many gravitinos would be produced thermally.
This mechanism might also allow $\phi$ to decay into
gravitinos. How our mechanism is related to an inflaton decay 
through supergravity effects 
\cite{gravitino}
merits further investigation.
Both of these effects
should provide non-trivial constraints on $f(\phi)$
from the cosmological data.

\begin{acknowledgments}
We would like to thank W. Fischler, R. Flauger, T. Futamase, P. B. Greene, N. Kaloper, M. Musso, M. Yamaguchi, and J. Yokoyama  for comments on the paper. E.K. acknowledges support from the Alfred P. Sloan Foundation.
\end{acknowledgments}

%%%%%%%%%%%%%%%%%%%%%

\end{document}